\documentclass[epj]{svjour}
\usepackage{graphicx}
\usepackage{grffile}
\usepackage{color}
\usepackage{amsmath}
\usepackage{ amssymb }
\usepackage{ esint }
\usepackage{cite}

\newcommand{\vek}[1]{{\mathbf{#1}}}

\definecolor{darkgreen}{RGB}{78,154,6}

\begin{document}

\title{Active Brownian particles moving in a random Lorentz gas}

\author{Maria Zeitz\inst{1}\thanks{\emph{Email address:} maria.zeitz@tu-berlin.de}, Katrin Wolff\inst{1},%
\and Holger Stark\inst{1}\thanks{\emph{Email address:} holger.stark@tu-berlin.de}
}                     
\institute{Institut f\"ur Theoretische Physik, Technische Universit\"at Berlin, Hardenbergstra\ss e 36, 10623 Berlin, Germany 
}

\date{Received: date / Revised version: date}

\abstract{
Biological microswimmers often inhabit  
a
porous or crowded 
environment
such as soil. In order to understand 
how such a
complex environment
influences their spreading,
we numerically study non-interacting active Brownian particles (ABPs) in 
a 
two-dimensional
random Lorentz gas.
Close to the percolation transition in the Lorentz gas, they perform the same subdiffusive motion as ballistic and diffusive particles.
However, due to their persistent motion they reach their long-time dynamics faster than passive particles and also show
superdiffusive motion at intermediate times. While above the critical obstacle density $\eta_c$ the ABPs are trapped, their long-time 
diffusion below $\eta_c$ is strongly influenced by the propulsion speed $v_0$. With increasing $v_0$, ABPs are stuck at the obstacles
for longer times. Thus, for large propulsion speed, the long-time diffusion constant decreases more strongly in a denser obstacle 
environment than for passive particles. This agrees with  the behavior of an effective swimming velocity and persistence time,
which we extract from the velocity autocorrelation function.
\PACS{
      {47.63.Gd}{	Swimming microorganisms}  \and
      {83.10.M}{	Molecular dynamics, Brownian dynamics}
       \and
      {87.10.Mn}{Stochastic modeling}
     } 
} 
\maketitle
\section{Introduction}
\label{intro}
Active matter 
has been
in the focus of 
intense
research 
for the last decade
 \cite{Ramaswamy2010,Romanczuk2012,Marchetti2013,Zottl2016}. 
On small 
length scales it 
describes the motion of biological microswimmers
such as
swimming bacteria or motile cells, and of artificial microswimmers, such as 
active
Janus particles or self-propelled 
emulsion
droplets \cite{Polin2009,Alizadehrad2015,Schmitt2013,Adhyapak2015,Schmitt2016,Maass2016}. 
In 
a homogeneous environment, microswimmers at low Rey\-nolds number 
first move ballistically and then cross over
to enhanced diffusion 
due to random rotational motion of their swimming direction
\cite{Zottl2016,Howse2007}.
Active Brownian particles 
provide a 
simple
stochastic model
for microswimmer such as active
Janus particles \cite{Romanczuk2012},
which in contrast to bacteria do not tumble.

Natural microswimmers usually do not move in homogeneous environments but encounter soft 
and
solid walls,
obstacles \cite{Drescher2011,Berke2008,Schaar2015,Takagi2014,Kaiser2012}, or even more complex environments like 
the intestinal tract 
\cite{Berg},
porous soil \cite{Ford2007}, 
and blood flow \cite{Engstler2007}.
A heterogeneous environment can be realized in different ways,
both in experiments and theory,
\emph{e.g.}, 
by regular 
or irregular
patterns of obstacles 
\cite{Park2008,Holm2011,Volpe2011,Heddergott2012,Majmudar2012,Johari2013,Chepizhko2013a,Chepizhko2013,Reichhardt2014,Schirmacher2015,Chepizhko2015b,Raatz2015,Munch2016},
mazes \cite{Khatami2016},
arrays of funnels \cite{Tailleur2009,Kaiser2012,Wan2008,Volpe2014,Galajda2007,Reichhardt2016}, pinning substrates \cite{Sandor2016a}, or patterned light fields, which control the velocity of the microswimmer
\cite{Volpe2014a,Pototsky2013}.
For a review see \cite{Zottl2016,Bechinger2016}.

In real systems the heterogeneities of the environment are mosty irregular.
One way to model them is the
random Lorentz gas \cite{Weijland1968a,Hofling2008b,Sch??bl2014}. 
In this approach the obstacles are fixed and randomly distributed with a given area fraction. 
The properties of the Lorenzt gas change fundamentally with varying density.
In particular, it shows a transition to
continuum percolation \cite{Mertens2012},
where the complementary free space stops to percolate through the whole system.
As a consequence, 
the dynamics of a test particle in such a Lorentz gas crucially depends on 
density. Above a critical obstacle density,
the
test particle 
can no longer
explore the whole environment 
but stays trapped
and long-range
transport is effectively suppressed. 
H\"ofling \emph{et al.} have observed that 
in two dimensions
and close to the critical density
diffusive and ballistic particles show the same 
subdiffusive motion
in a random Lorentz gas
\cite{Hofling2008b,Spanner2011,Bauer2010,Spanner2016,Schnyder2015}. 

Microswimmers can interact in different ways with obstacles. 
For example, Chepizhko \emph{et al.} have studied self-propelled particles in an environment of randomly distributed 
point-like obstacles.
Instead of implementing
steric 
interactions,
they let the self-propelled particles deflect from the point-like
obstacles with a characteristic turning speed \cite{Chepizhko2013a}. Depending
on the system parameters,
this setting 
can lead to trapped states where particles circle around an obstacle
and thus to 
subdiffusive motion.
It also shows interesting effects 
during
collective motion.
For example, transport in the presence of obstacles is optimized by a specific noise value
\cite{Chepizhko2013,Chepizhko2015b}. 

In this article we will investigate an active Brownian particle (ABP) moving in a random Lorentz gas
implementing explicit steric interactions between ABP and obstacles.
We will demonstrate 
that it performs the same subdiffusive motion as ballistic and diffusive particles close to percolation. Besides this universal feature,
ABPs  explore their environment faster than passive particles due to their persistent motion and therefore reach their 
long-time dynamics at earlier times.  At intermediate times their dynamics is superdiffusive. A determining characteristic of ABPs is that 
they are stuck to the obstacles due to their self-propulsion. 
This has consequences for the diffusive spreading below the critical obstacle density. Namely,
for large propulsion speed, the long-time diffusion constant 
decreases more strongly in a denser obstacle environment than for passive particles. We rationalize this behavior by 
studying the velocity autocorrelaton function, which motivates us to introduce an effective swimming velocity and persistence time.

The article is structured as follows. In sect.\ \ref{sec:model}
we 
introduce the equations of motion 
of
the ABP and 
model 
the 
obstacle
environment as a random Lorentz gas.
In sect. \ref{sec:passive} we
recap the 
dynamics of a
passive Brownian particle in 
the Lorentz gas and compare it
to the 
ABP
in sect. \ref{sec:ABP}. 
Finally, in
sect. \ref{sec.same_random} we discuss how
the long-time diffusion coefficient of an ABP and its
persistent motion
is reduced in the presence of obstacles
using the velocity autocorrelation function. We
finish with concluding remarks in sect. \ref{sec:summary}.

\section{Introduction of the model}
\label{sec:model}

\subsection{Active Brownian particle interacting with obstacles}

Our model of a microswimmer is a  
circular active Brownian particle (ABP) with radius 
$R_s$.
It moves with a constant velocity $v_0$ along a direction $\vek e$ in a 
two-dimensional 
environment
filled with circular obstacles of radius $R_o$.
The particle
also
experiences rotational and translational 
thermal noise.
Thus, the
Langevin equations 
governing the dynamics of the ABP reads \cite{Romanczuk2012,Zottl2016}
\begin{align}
\begin{split}
\frac{\mathrm d \mathbf{ r} (t)}{\mathrm d t }&=v_{0}\mathbf e (t)+\mu^{T}\sum_{i}\mathbf F_{o}^{i}+\sqrt{2D^{T}}\boldsymbol\eta^{T}(t), \\
\frac{\mathrm d \mathbf e(t)}{\mathrm d t }&=\sqrt{2D^{R}}\boldsymbol\eta^{R}(t)\times \mathbf e(t) \, ,
\end{split}
\label{eq:langevin}
\end{align}
where $D^T$ and $D^R$ are the respective translational and rotational diffusion constants. Furthermore,
$\boldsymbol\eta^{T}(t)$ is a rescaled stochastic force and $\boldsymbol\eta^{R}(t)$ 
a rescaled stochastic torque, both with 
zero mean and
Gaussian white noise correlations:
\begin{align}
	\langle\boldsymbol\eta^{T,R}(t)\rangle&=0,\label{eq:noise_mean}\\
	\langle\boldsymbol\eta^{T,R}(t_{1})\otimes\boldsymbol\eta^{T,R}(t_{2})\rangle&=\mathbf 1 \delta(t_{1}-t_{2})\label{eq:noise_corr}.
\end{align}
Note that 
the stochastic torque 
always
points
out of
the $x,y$
plane, $\boldsymbol \eta ^R = \eta_z^R \mathbf e_z$.

Finally, the
interaction
force $\mathbf F_{o}^{i}$ between an ABP and obstacle $i$ is modeled by
pure volume exclusion.
It
derives from the
\textit{Weeks-Chandler-Anderson} potential, which is 
 a 
 Lennard-Jones potential 
 cut off at the minimum:
\begin{align}
V=\begin{cases}
4\varepsilon\left[\left(\frac{\sigma}{r}\right)^{12}-\left(\frac{\sigma}{r}\right)^{6}\right] +\varepsilon &\text{for } r<d^{*}\\
0&\text{for } r\geq d^{*} \, .
\end{cases}
\label{eq:potential}
\end{align}
Here,
$d^{*}=2^{1/6}\sigma=R_s+R_o$
is the distance, where the potential is minimal and 
where it also
vanishes.
The resulting force in Eq. \eqref{eq:langevin} is then given by $\mathbf{F}_o =- \boldsymbol{\nabla} V$.
Hydrodynamic interactions 
are neglected in this model.

The motion of the active Brownian particle in free space is fully characterized by the dimensionless P\'{e}clet number, 
which compares the times needed for diffusive 
and
active motion along a distance $R_s$, 
\begin{equation}
\text{Pe}=\frac{2R_s v_0}{D^T} \, .
\label{eq.Pe}
\end{equation}
Moreover, in a homogeneous bulk system the 
ABP
moves ballistically on the
persistence length 
$ \Delta s _{r} = v_{0}\tau_{r}$,
where $ \tau_{r}=1/D^{R}$ is the persistence time 
completely determined by the rotational diffusion
constant.

\subsection{Lorentz model and continuum percolation}

\label{sec:model_env}

\begin{figure*}
\begin{center} 
	\resizebox{0.7\textwidth}{!}{%
	\includegraphics{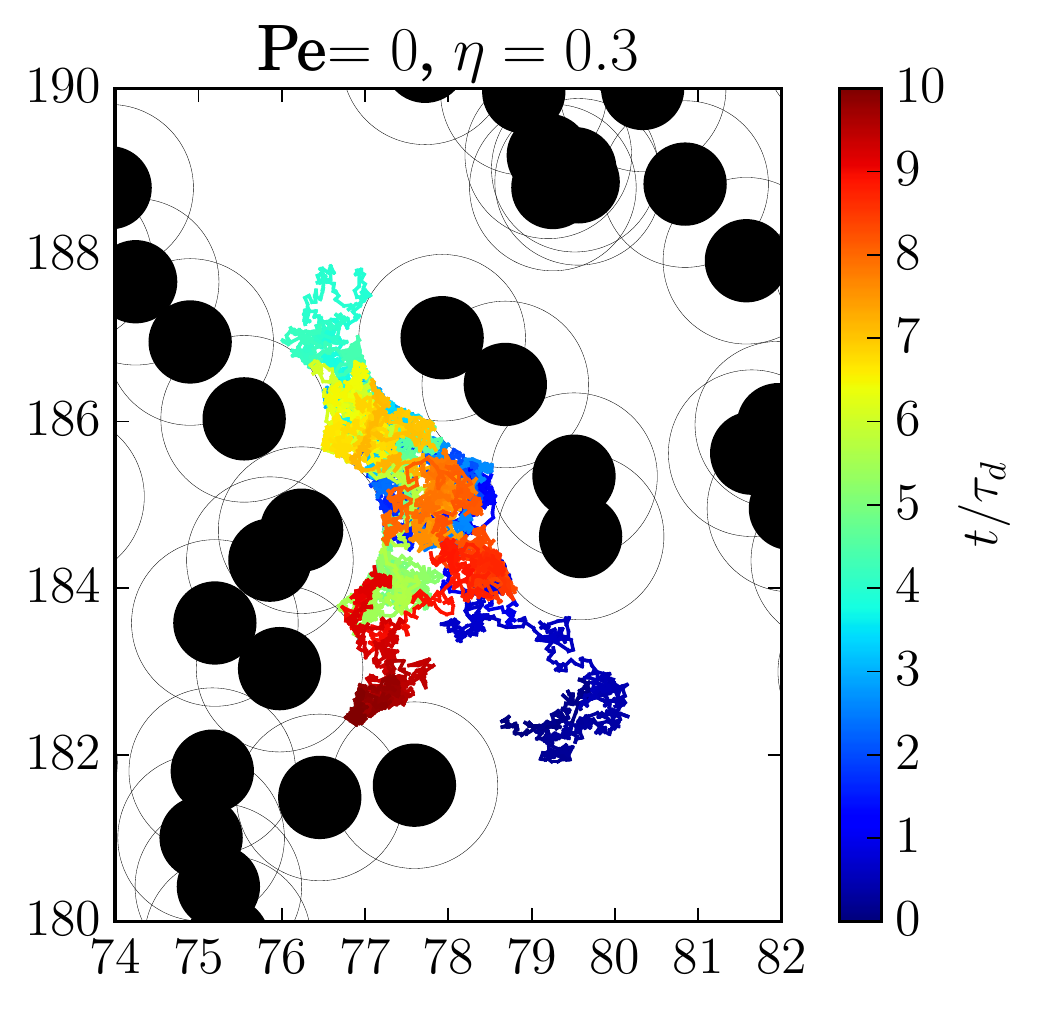}
	\includegraphics{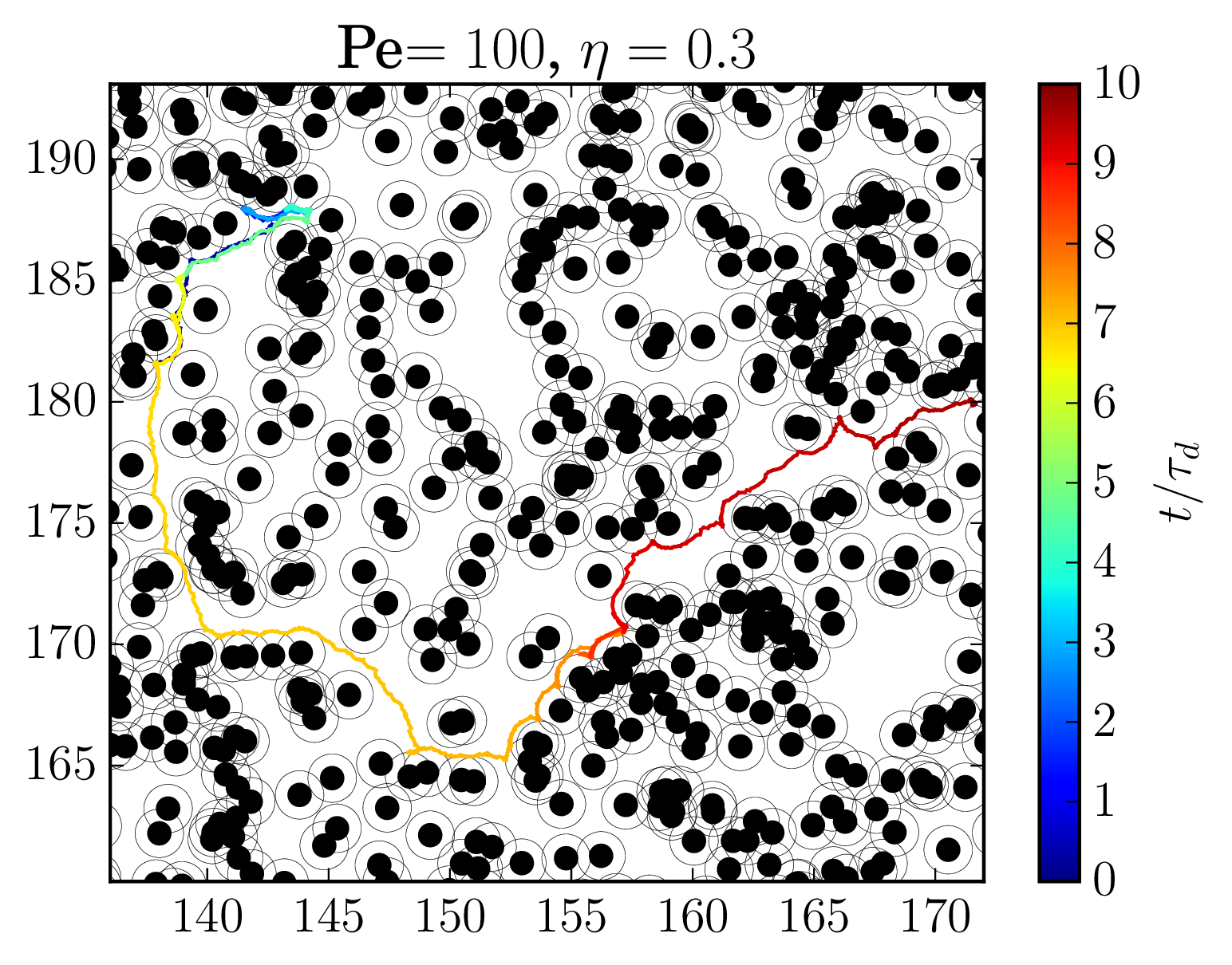}
	}
\end{center}
   \caption{Left: Trajectory of a passive Brownian particle in a heterogeneous environment with reduced obstacle density $\eta=0.3$. 
   Right: Trajectory of an 
   ABP with $\mathrm{Pe} = 100$ under the same conditions. Time along the trajectories is encoded by color 
    according to the color bar.
}
   \label{fig:trajectory}
\end{figure*}

We model the 
heterogeneous
environment 
by a random Lorentz gas, where 
circular
obstacles are randomly distributed in the plane 
and
with positions fixed in space
(spatial Poisson point process)
 \cite{Stauffer1994}.
The obstacles are assigned a
reduced density 
or area fraction
$\eta=  \pi R_{o}^2/A$,
where $A$ is the area of the system. However, since the
obstacles can fully penetrate each other, 
the actual area fraction 
is \cite{Torquato2002}
\begin{equation}
\phi_o=1-e^{-\eta} \, .
\label{eq:phio}
\end{equation}

Percolation theory gives some first insights 
how an ABP moves through
the random Lorentz gas.
It provides a critical density for the transition to an infinite cluster of overlapping disks percolating through the whole system
\cite{Torquato2002} or for the percolating free space between obstacles
\cite{Bauer2010}.
In two dimensions the critical particle area fraction for the percolating
cluster of connected disks is $\phi_{o,c}=0.67637\pm0.00005$ \cite{Torquato2002}, while 
the critical void 
area fraction for the percolating free space
is $\phi_{v,c}\approx0.324$ \cite{Bauer2010}.
The latter determines whether the ABP ultimately performs diffusive motion or becomes localized (localization transition). In two dimensions the
two area fractions add up to 
one,
because 
the percolating clusters of free space and of overlapping disks are complementary to each other. They can only coexist
in strongly anisotropic systems but not in the random Lorentz gas with its isotropic distribution of disks. In three dimensions this
restriction is no longer valid.

In our simulations we 
always
choose the reduced density $\eta$ instead of the actual area fraction $\phi_o$. 
Using Eq.\ \eqref{eq:phio} the critical value of the reduced obstacle density is
$\eta_c =  -\ln\left(1-\phi_{o,c}\right) 
\approx 1.12815$.
Now, the ABP
itself has a finite 
extent,
so the accessible space for 
its center of mass 
is determined not only by the obstacles but also by the swimmer radius.
The 
ABPs can only pass 
between two obstacles if their distance is larger than $2(R_{o}+R_{s})$ assuming ideal hard-core interactions.
Thus,
we can map the problem of an extended swimmer in a random environment onto a point-like swimmer in an environment of obstacles 
with an effective radius $R_\text{eff} = R_{o}+R_{s} = 2 R_{o}$
and an
effective reduced density 
$\eta^{\text{eff}}=N\pi(R_\text{eff})^{2}/V = 4 \eta$ 
(here and in the following we assume $R_{o}=R_{s}$).
Therefore, we
find for the critical reduced density of the actual obstacles
\begin{equation}
\eta_{c} = \eta_c^\text{eff} / 4 \approx 0.28 \, .
\end{equation}

We will see 
that in our system the localization transition takes place at densities larger than
the theoretically predicted value $\eta_c \approx 0.28$.
This is mainly due to a finite-size effect.
Near to the critical density, the 
size of the largest
finite cluster of free space (or the 
largest trap) diverges as
the correlation length: 
$\xi\sim|\eta-\eta_{c}|^{-\nu}$. 
The
exponent $\nu=4/3$
is
known from lattice percolation and also valid for continuum percolation.
In our simulations we can only realize a finite system with size $L$ and apply
periodic boundary conditions, in order to mimic infinite systems. 
The system size $L$ is then an upper bound
for
the size of the largest cluster of accessible space 
given by $\xi$.
Coming from high densities $\eta$, 
the free space starts to percolate for $\xi \approx L$
and the
apparent
critical density in a finite system is shifted 
according to
$\eta_{c}(L)-\eta_{c}(\infty)\sim L^{-1/\nu}$. 
In contrast to Ref.
\cite{Bauer2010},
where the system size was
 $L/R_o=10000$, 
we use much smaller systems  
with
a fixed number of $15000$ obstacles.
Depending on $\eta$, the system size 
becomes
$L = \sqrt{A} = \sqrt{15000\pi R_{o}^{2}/\eta}$,
which ranges
from 140 to 343
for $\eta \in [0.1,0.6]$ used in our simulations. This is a factor 100 smaller than in \cite{Bauer2010} and roughly gives a shift
of
$\eta_{c}(L)-\eta_{c}(\infty)\sim 0.02$, in agreement with our results presented below.
Finally, note that in contrast to typical studies of percolation, we keep
the number of obstacles fixed rather than the system size.
This makes sense since active particles strongly accumulate at bounding surfaces which, therefore, strongly influence the 
dynamics of ABPs \cite{Schaar2015}.

\subsection{System parameters}

For all
of our simulations we use an environment 
with
15000 obstacles. To determine the long-time behavior of the ABPs, we perform simulations with
500 non-interacting swimmers in the same environment.
In order to study the 
short-time
behavior and especially the velocity autocorrelation function in more 
detail, we simulate
100 non-interacting swimmers in upto 
six different realizations of the environment, so 600 trajectories in total.

We randomly place the obstacles on a square with area $A$ and allow them to overlap. Then, the swimmers are
randomly distributed but overlaps with obstacles are not allowed. If they occur, the relevant swimmers are newly placed.
Since ABPs eventually accumulate at the surfaces of the obstacles, the random swimmer distribution is not the steady state of
the system. So, before we start our data acquisition, we let the system equilibrate 
during
a time $0.5 \tau_d$.
Here,
\begin{equation}
\tau_d = (2 R_s)^2 / D^T
\label{eq.tau_d}
\end{equation}
is the time a passive particle needs to 
diffuse
its own size, while the ABP 
moves over
much longer distances during $\tau_d$ depending on the P\'eclet number.

In the following, we set $R_s=R_o = R$. We give lengths in units of $2R$ and rescale time by $\tau_d = (2 R)^2 / D^T$.
In three dimensions the respective thermal diffusion coefficients for translation and rotation are related by $D^R=3D^T/(4R^2)$ \cite{Zottl2016}.
Albeit in two dimensions,
we use the same ratio in all our simulations and the ratio of the diffusion times becomes $\tau_r / \tau_d = 1/3$.
Finally, for the time step $\Delta t$ in the Langevin-dynamics simulations we choose $\Delta t \leq 10^{-5}$ in units of $\tau_d$,
where we 
need
to adjust $\Delta t$ to smaller values with increasing P\'eclet number.

\section{Influence of the random environment on a passive Brownian particle}
\label{sec:passive}

Before presenting our results for ABPs, we review some basic results 
for
passive Brownian particles. 
Figure\ \ref{fig:trajectory} shows a close-up of the system with obstacles in black. 
The
surrounding excluded volume due to the finite  size of the Brownian particles is indicated 
by gray rings. 
We show the
trajectories of the passive and active Brownian particles,
where 
color encodes time.
The trajectory of the center of the passive particle (left) is only governed  by diffusion. 
It explores the accessible free space and moves about eight particle diameters away from the 
starting point as expected for $t=10$.

\begin{figure*}
\begin{center} 
	\resizebox{\textwidth}{!}{%
	\includegraphics{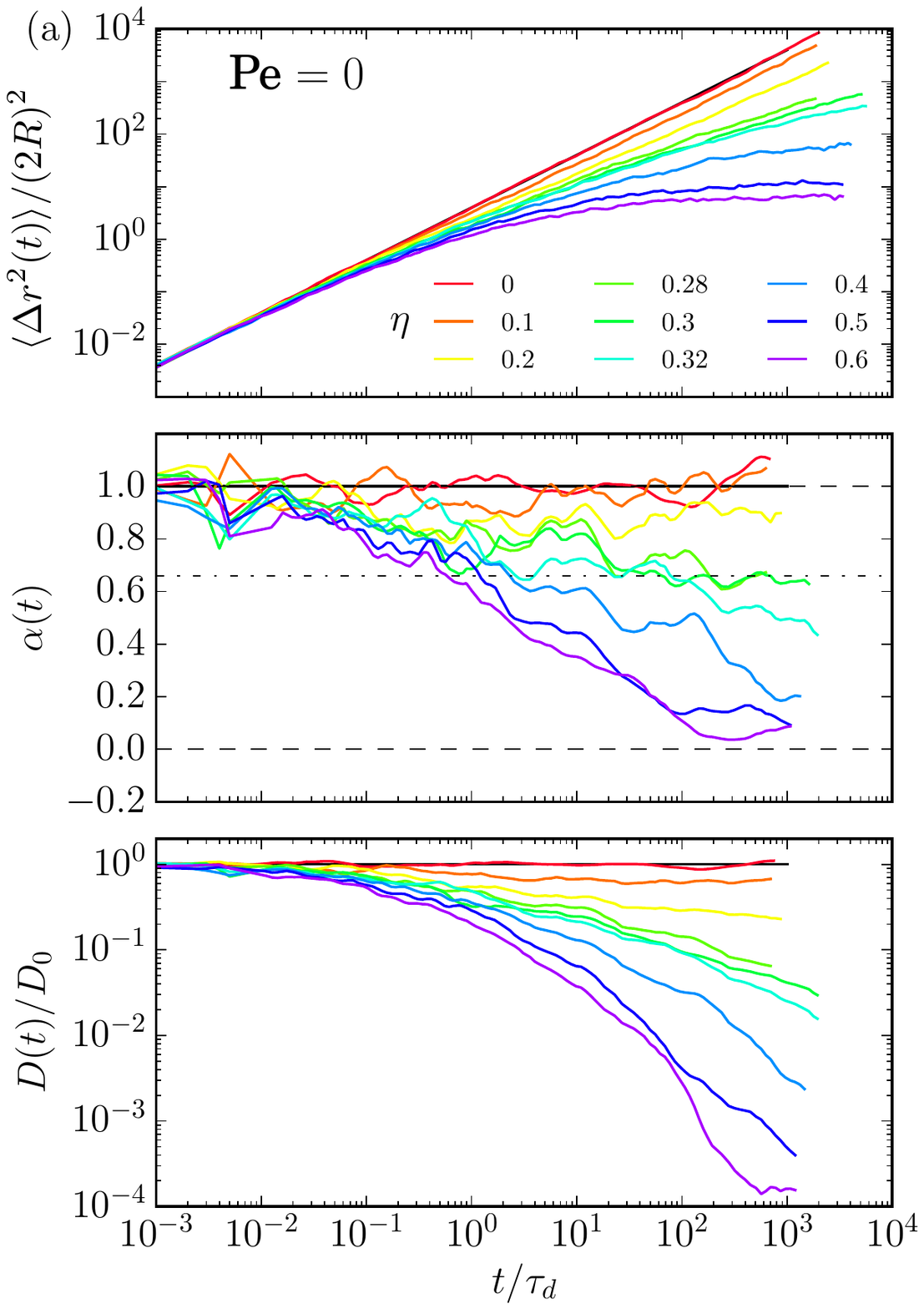}
	\includegraphics{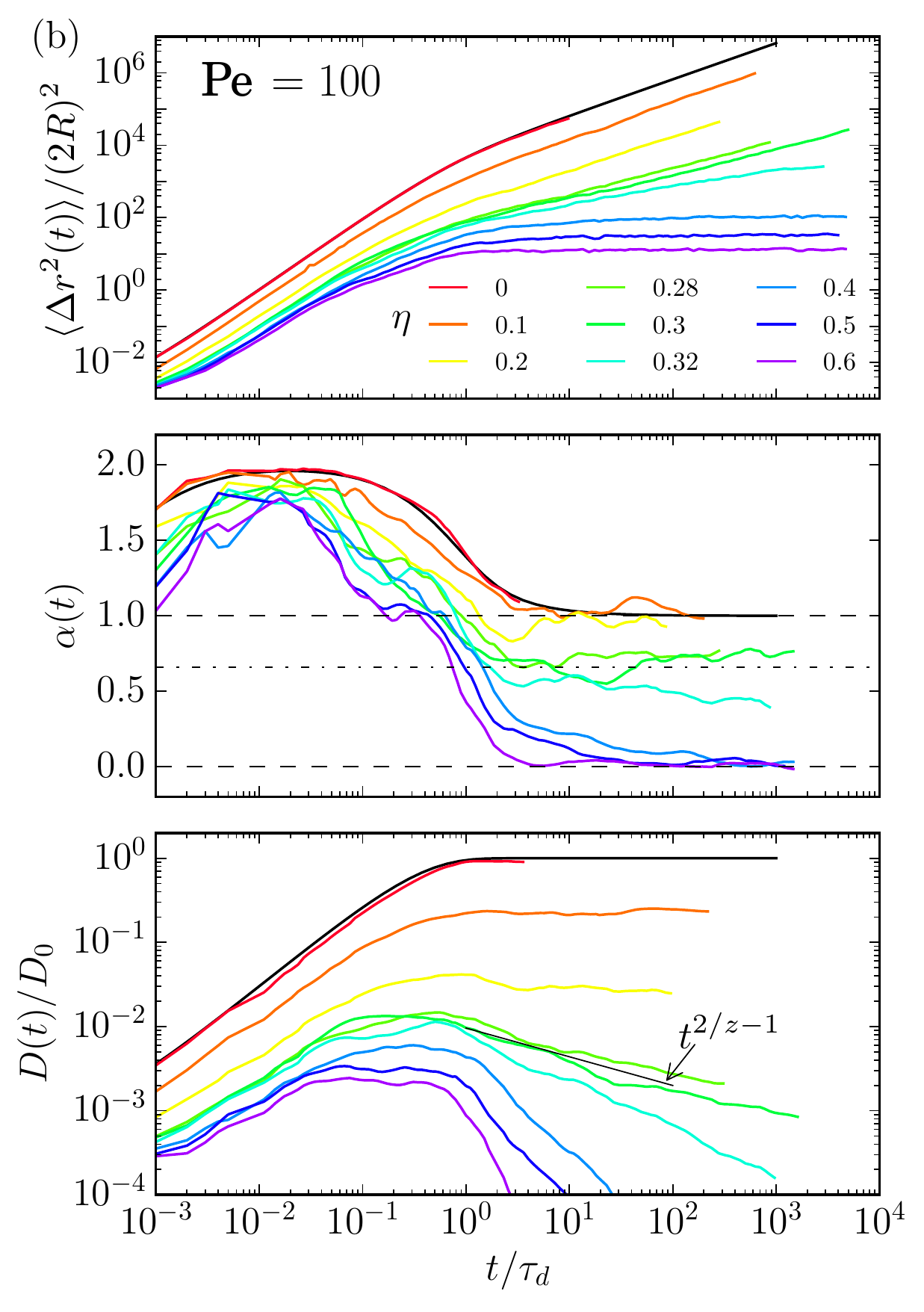}
	}
\end{center}
   \caption{Mean squared displacement $\langle \Delta r ^{2}  \rangle $ (top), local exponent $ \alpha (t) $ 
   (center),
    and local diffusion coefficient $ D (t) / D_{0}$ 
    (bottom)
    for (a) a passive and (b) an active Brownian particle. 
    The dashed-dotted lines in the middle graphs indicate $\alpha = 2/z \approx 0.66$.
}
   \label{fig:msd_0_100}
\end{figure*}

To be more quantitative,
we
discuss
the mean squared displacement, the local exponent $\alpha(t)$, and the local diffusion coefficient $D(t)$ 
for different reduced obstacle densities
as shown in Fig. \ref{fig:msd_0_100}(a).

\subsection{Mean squared displacement}
For a 
Brownian
particle in two dimensions,
diffusing in free space, the mean squared displacement is given by $ \langle \Delta r^2(t) \rangle = 4 D^T t$ and shown 
by
the black curve in the upper graph of
Fig.\ \ref{fig:msd_0_100}(a).
At small obstacle densities
($\eta = 0.1$, 0.2)
the mean squared displacement 
behaves similar
to the free case, however, with a 
smaller
diffusion coefficient 
(note the double-logarithmic plot).
For very high obstacle densities 
($\eta = 0.5$, 0.6)
it
eventually saturates at finite values. At intermediate obstacle densities we observe subdiffusive behavior 
also at long times.
Thus, the mean squared displacement shows a
transition 
from a linear to a localized regime via subdiffusion. 
In an inifinitely extended system, the
latter occurs exactly at the critical density $\eta_c$, where
the free space does not percolate any more.
Beyond this density
particles are trapped in 
regions of
finite size
and the mean squared displacement 
saturates at 
$ \langle \Delta r^2(t) \rangle$ equal to
the areas of these regions. All this is thoroughly reviewed in Ref.\ \cite{Bauer2010}.
We will now be more quantitative to be able to compare with the ABPs.

\subsection{Local exponent} \label{subsec.exp}

Subdiffusive behavior is indicated by $ \langle \Delta r^2(t) \rangle \propto t^{\alpha}$ with an exponent $\alpha < 1$. Thus we
determine
the local exponent 
\begin{equation}
\alpha(t)=\frac{\mathrm d  \log\langle \Delta r^2 (t)\rangle}
{\mathrm d  \log t } \, .
\end{equation}
from the mean squared displacement and plot it in the middle graph of Fig.\ \ref{fig:msd_0_100}(a).
In free space
($\eta = 0$)
$\alpha(t)$ does not change in time and is always one. Also for low $\eta$ the exponent 
fluctuates around one,
which indicates that the particle dynamics is still governed by 
conventional
diffusion and the obstacles do not have much impact,
whereas for very high $\eta$ the exponent 
tends
to zero 
as expected when particles become trapped or localized.
For intermediate obstacle densities $\eta$ the local exponent 
$\alpha(t)$ 
is smaller than one and indicates
subdiffusive behavior. 
In particular, at $\eta = 0.28$ the exponent reaches the value $\alpha = 2/z = 2/3.036 \approx 0.66$
as expected at the percolation transition and shown by the dashed-dotted line.
However, for most parameters the exponent has not yet reached a stationary value.

In an inifinite system close to percolation, particles exhibit subdiffusion below the correlation length $\xi$
of the fractal cluster of free space. Then,
on lengths larger than $\xi$ 
they
either become trapped ($\eta < \eta_c$) or diffusive ($\eta > \eta_c$) \cite{Torquato2002}. 
Subdiffusion persists exactly at $\eta = \eta_c$, where $\xi$ diverges. In finite systems the correlation length $\xi$ is bounded by the 
system size $L$. So any subdiffusive
motion at short times with $\langle \Delta r^2(t) \rangle < \xi ^2 $ is only transient and will ultimately become trapped for $\eta \ge \eta_c(L)$
with an exponent $\alpha = 0$ or turn into normal diffusion for $\eta < \eta_c$ with an exponent $\alpha = 1$.
Thus, in Fig.\ \ref{fig:msd_0_100}(a), middle graph the curve for $\eta = 0.28$ and probably also for $\eta = 0.3$ will ultimately 
reach one for long times beyond our simulated times, whereas the curves for larger $\eta$ clearly tend towards zero.

Passive particles diffuse slowly and in contrast to active particles need a long time to explore their environment. 
Together with the diverging $\xi$ close to the percolation transition, this explains why in
Fig. \ref{fig:msd_0_100} only the exponents far from the critical density 
($\eta = 0.1, 02$ and $\eta = 0.6$) have reached 
the
stationary values of one and zero, respectively. 

\subsection{Local diffusion coefficient}

Another quantity to characterize the mean squared displacement is the
local diffusion coefficient
\begin{equation}
D(t) = \frac{1}{4} \frac{\mathrm{d}\langle \Delta r^2(t) \rangle}{\mathrm{d} t} \,.
\label{eq:diffusion}
\end{equation}
In the lower graph of Fig.\ \ref{fig:msd_0_100}(a) we plot it normalized by the
diffusion coefficient in free space, $D_0$, which is $D^{T}$ for passive particles. 
Already at low densities $\eta$ the diffusion coefficient $D(t)$ is not constant in time. 
At
$\eta=0.1$
it slightly decreases and around $t=\tau_d$
saturates 
at a
ratio $D(t) / D_0$ below one.
For 
$\eta = 0.2$ the ratio decreases further and
takes more time 
to become stationary. 
For 
obstacle densities above the 
percolation
transition $D(t)$ 
tends
to zero
as expected.
In agreement with the discussion in Sec.\ \ref{subsec.exp},
 at intermediate densities the time to reach a diffusive state with a constant $D(t)$ or a localized state with $ D(t)=0 $ 
becomes very large 
close to
the critical density $\eta = \eta_c(L)$.
 
Analyzing the local exponent $\alpha (t)$ and the local diffusion coefficient $ D(t) / D_{0}$
shows that
obstacles 
at low densities $ \eta $ act as an additional source of noise,
while at high obstacle densities $ \eta $ they confine the available space of the swimmers to a finite extent.
A good 
discussion
of the scaling behavior of the dynamics of Brownian particles in a heterogeneous environment close to percolation
is found in \cite{Bauer2010}.

\section{Influence of the random environment on a microswimmer}
\label{sec:ABP}

In this section we 
first
study how the complex environment influences the motion of 
ABPs moving at $\mathrm{Pe}=100$, as an example. We 
show
how the dynamics of the swimmer changes with obstacle density. The results to be discussed below are plotted in 
Fig.\ \ref{fig:msd_0_100}(b). In 
sect.\ \ref{subsec.percol}
we will compare ABPs of different velocities 
close to the percolation transition.

In contrast to a passive particle the ABP is much more in contact with the obstacles
and covers a longer path due to its swimming velocity. This is demonstrated in Fig.\ \ref{fig:trajectory}(b)
for
an example trajectory 
at
$\mathrm{Pe}=100$.
The ABP spends less time in the free space but is rather guided along the walls of obstacle clusters. 
Due to its overdamped motion the ABP just stops if its self-propulsive velocity points perpendicular to a bounding
wall. However, as
long as the swimming direction has a component parallel to 
an
obstacle surface, it will slide along the surface. 
The ABP leaves the obstacle 
due to rotational diffusion of its swimming direction (governed by the persistence time $\tau_r$) and because 
the 
obstacle
is curved.

\begin{figure}
\begin{center} 
	\resizebox{0.50\textwidth}{!}{%
	\includegraphics{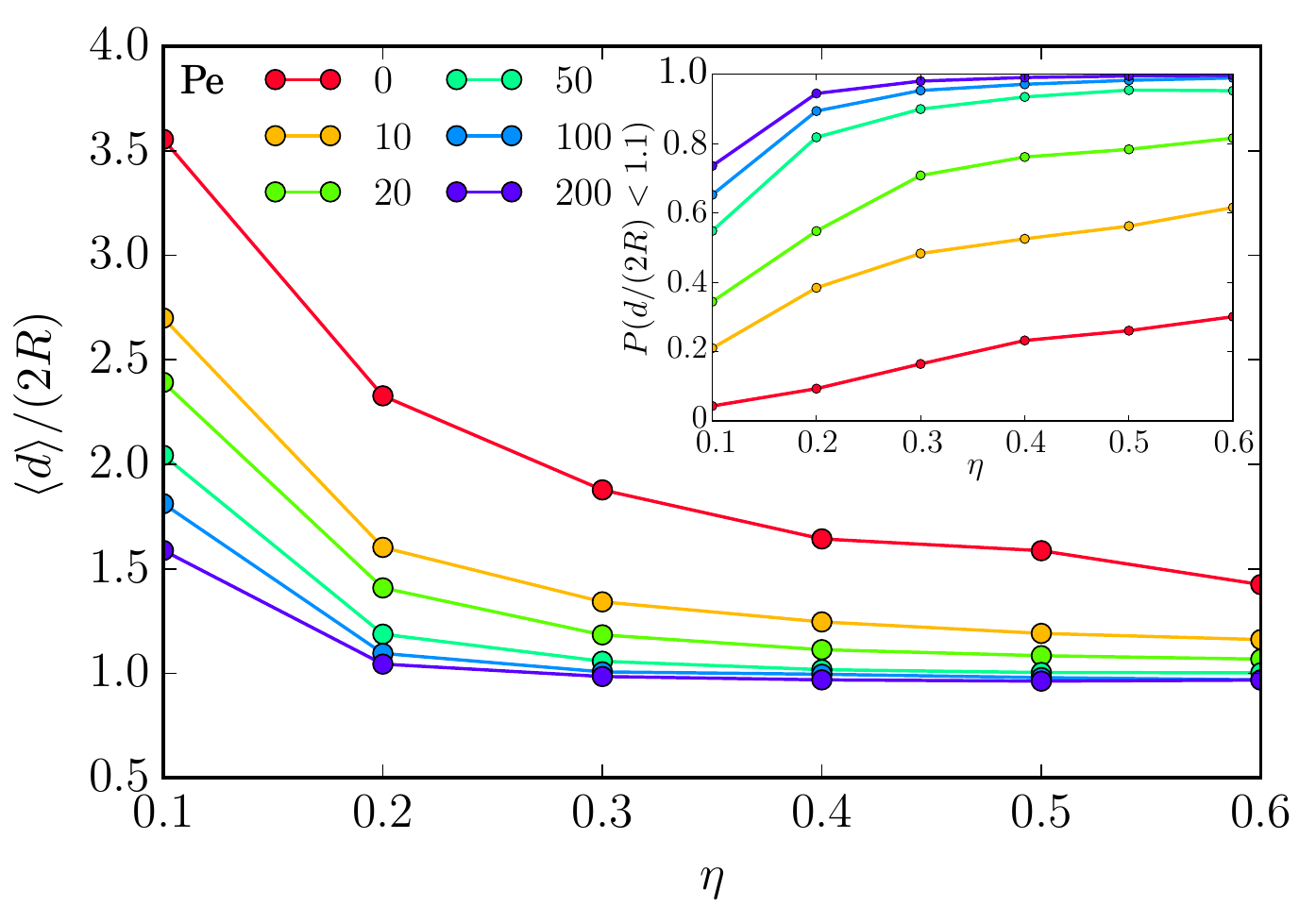}}
\end{center}
   \caption{Mean distance of an ABP to the
    closest obstacle, $\langle d \rangle / (2R)$, plotted versus $\eta$ for different $\mathrm{Pe}$.
    Inset: Probability for the ABP to have a distance $d$ smaller than $1.1 (2 R)$ to the nearest obstacle,  $P(d /(2R) < 1.1)$.
}
   \label{fig:mean_distance}
\end{figure}

Indeed, Fig.\ \ref{fig:mean_distance} demonstrates that the mean distance of an ABP 
to
the closest obstacle drastically decreases with increasing P\'eclet number at the same obstacle density $\eta$. 
In partiular, for $\mathrm{Pe} \geq 50$ the mean distance is close to 
$2 R$. It varies only little with increasing $\eta$ and all ABPs reside close to or at obstacle surfaces. Correspondingly, 
the probability of an ABP to be at a distance smaller than $1.1(2R)$
to the nearest obstacle increases with $\mathrm{Pe}$ and approaches one (see inset of Fig.\ \ref{fig:mean_distance}).

\subsection{Mean squared displacement}
\paragraph{Free space}
For 
an ABP moving
with velocity $v_0$ the mean squared displacement in free space can be written as
\cite{Howse2007,Downton2009,TenHagen2011}:
\begin{equation} 
	\langle\Delta r^{2}(t)\rangle=4D^{T} t+2v_{0}^{2}\tau_{r}t-2v_{0}^{2}\tau_{r}^{2}\left(1-e^{-t/\tau_{r}}\right).
	\label{eq:msd}
\end{equation}
From this equation 
one infers
two relevant time scales and an effective diffusion coefficient 
at
long times. On times
shorter than 
\begin{equation}
	\tau_{1}=4D^{T}/v_{0}^{2} = 4 \tau_d / \mathrm{Pe}^2 \, ,
	\label{eq:tau1}
\end{equation}
where we used Eqs.\ (\ref{eq.Pe}) and (\ref{eq.tau_d}) to derive the second expression,
translational diffusion 
dominates over
active propulsion and the 
ABPs move by thermal diffusion.
Typically, this
regime is only observed 
at
small P\'eclet numbers. On times
up to the persistence time $\tau_{r}=1/D^{R}$
(in two dimensions), 
self-propulsion dominates and 
the ABPs
move ballistically. On 
times larger than $\tau_r$ the particle orientation decorrelates and
diffusive motion
occurs
with an increased effective diffusion coefficient $D_\text{eff}=D^{T}+ 1/2 v_{0}^{2} \tau_{r}$.
The ballistic regime and the crossover  to enhanced diffusion around $\tau_r = \tau_d / 3$ is clearly illustrated
in the upper graph of Fig.\ \ref{fig:msd_0_100}(b). The initial diffusive regime occurs at times smaller than 
$\tau_1 = 4 \cdot 10^{-4} \tau_d$ and is not visible in the graph.

\paragraph{Random environment}
Here the mean squared displacement in Fig.\ \ref{fig:msd_0_100}(b) follows the general trend set by passive particles and ABPs
in free space. An inital ballistic or superdiffusive regime is followed by either effective diffusion at small densities $\eta$
or a localization at high $\eta$, where  $\langle \Delta r^{2} (t) \rangle $ saturates at the square of the mean trap size.
Close to the critical percolation density
we now observe a subdiffusive regime, which extends over several decades in time
compared to the passive case.
The time spent in the subdiffusive regime becomes shorter further away from the critical density
(see for example $\eta = 0.4$).

Due to their
self-propulsion, 
ABPs cover 
a much longer
distance 
than passive particles
in the same time 
(see 
Fig.\ \ref{fig:trajectory}).
As a result, they explore the confining space of a trap much faster. Thus, the mean squared displacement of ABPs 
at densities $\eta = 0.4$ and higher
saturates at a pronounced plateau
much earlier than for passive particles.

At short times, passive particles or slow swimmers explore the whole free space by diffusion with the same diffusion coefficient.
Therefore, all the curves for different $\eta$ in the upper graph of Fig.\ \ref{fig:msd_0_100}(a) initially lie on top of each other.
In contrast, for fast ABPs the inital ballistic/superdiffusive motion slows down with increasing $\eta$. ABPs strongly accumulate
at the obstacles and the time they spend in free space decreases with $\eta$. Thus, effectively also their mean motility decreases 
with increasing $\eta$ at short times, which explains that the mean squared displacement in Fig.\ \ref{fig:msd_0_100}(b) is shifted 
downwards. 
In sect.\ \ref{sec:VACF} we will introduce an effective swimming velocity, which rationalizes the observed behavior.

\subsection{Local exponent}

\paragraph{Free space}
In the middle graph of Fig.\ \ref{fig:msd_0_100}(b)
the black solid line shows the analytic result for the local exponent of an 
ABP moving
in free space, which we directly calculate using Eq.\ (\ref{eq:msd}):
\begin{equation} 
	\alpha (t)  =\frac{ \mathrm d \log \langle \Delta r ^{2} \rangle  }{\mathrm d \log t } =
 	\frac{ 4 D^{T} t + 2 v_{0}^{2} \tau_{r} t - 2 v_{0}^{2} \tau_{r} t e^{  - t / \tau_{r} } }{\langle \Delta r ^{2} \rangle} \, .  
	\label{eq:alpha_free}
\end{equation}
Rescaling times by $\tau_d$ and lengths by the diameter $2R$, one shows that
the temporal evolution of the local exponent 
only
depends 
on the P\'eclet number, which also sets the time scale for the
transition from diffusion to ballistic movement, as $\tau_1$ in Eq.\ (\ref{eq:tau1}) demonstrates.
The second time scale $\tau_d$
marks the transition from ballistic back to diffusive 
motion.

\paragraph{Random environment}
In contrast to passive particles, the ABPs reach their long-time behavior much faster at times around $\tau_d$
and the transition is more pronounced. This is clearly demonstrated by the graphs for $\alpha(t)$ in Fig.\ \ref{fig:msd_0_100}.
The exponent 
of
the ABP assumes $\alpha=1$ for the diffusive motion at small $\eta=0.1, 0.2$, becomes zero at high obstacle
densities $\eta = 0.4, 0.5, 0.6$, and indicates clear sub\-diffusive motion at the intermediate densities $\eta=0.28, 0.3$.
As for passive particles we roughly find $\alpha$ close to 0.66 (dashed-dotted line) indicating that subdiffusion close to
the critical percolation density is mainly governed by the topology of free space and not by the difference between Brownian and 
ballistic active motion.

At short times
the ballistic/superdiffusive regime shrinks in the presence of obstacles
with increasing $\eta$ and also the local exponent $\alpha$ decreases. In particular,
the crossover time from ballistic/superdiffusive motion to diffusion shifts to smaller times
when $\eta$ becomes larger. 
Interestingly, at
high obstacle densities, $\eta \geq 0.4$,
a short
diffusive regime with $\alpha=1$ is established
before 
localization
with $\alpha = 0$. 
occurs.

\subsection{Local diffusion coefficient}
\paragraph{Free space}
We readily calculate the local diffusion coeffcient of an ABP using Eq.\ (\ref{eq:msd}):

 \begin{equation} 
 D(t) =  \frac 1 4\frac {\mathrm{d}\langle \Delta r^2(t) \rangle}{\mathrm{d}t} = D^{T} + \frac 1 2 v_{0}^{2} \tau_{r} - \frac 1 2 v_{0} \tau _{r} e^{- t / \tau_{r} }  \, .
 \label{eq:D-free}
 \end{equation}
It is plotted as the solid black line in the bottom graph of Fig. \ref{fig:msd_0_100} and shows a plateau at the
effective diffusion coefficient 
$D_\text{eff}=  D^{T} +  v_{0}^{2} \tau_{r}/2$, as already discussed.

\paragraph{Random environment}
Very prominently, at low $\eta$ the steady-state diffusion coefficient strongly decreases from the free-space value. Thus, even
at low densities, the obstacles 
have a much more pronounced effect compared to passive particles and strongly decrease the mobility of ABPs.
For 
high obstacle densities, $\eta \geq 0.4$,
one clearly observes how
$D(t)$ 
approaches
zero.
This occurs at smaller times, when $\eta$ increases.
Finally,
in the vicinity of the percolation transition the diffusion coefficient 
behaves close to
the universal power law, $D(t) \propto t^{2/z-1}$, as indicated in the graph.

\subsection{Microswimmers in a random Lorentz gas close to the percolation transition}
\label{subsec.percol}

\begin{figure}
\begin{center} 
	\resizebox{0.50\textwidth}{!}{%
	\includegraphics{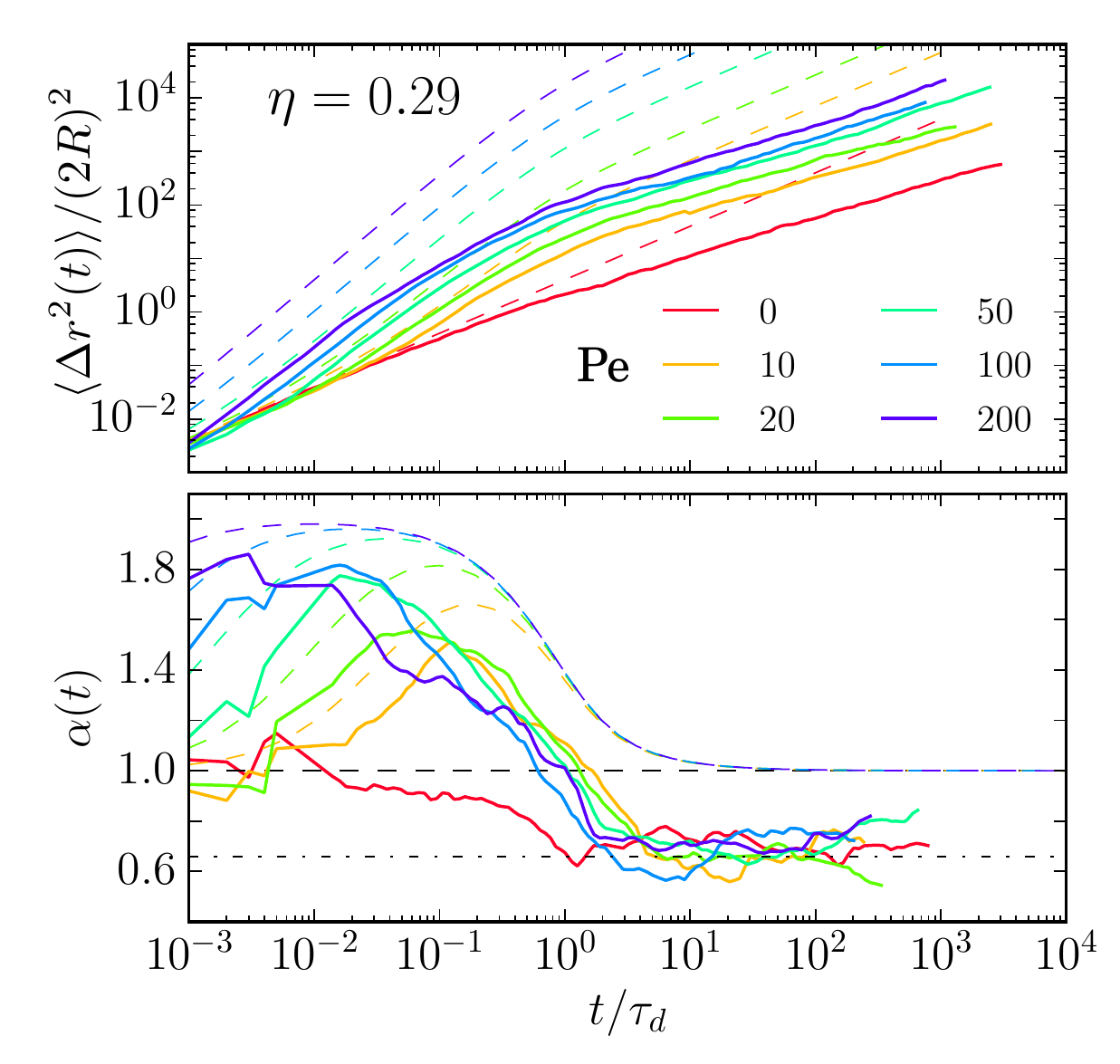}}
\end{center}
   \caption{Mean squared displacement $\langle \Delta r ^{2}  \rangle $ (top) and local exponent $ \alpha (t)$ 
(bottom)
   for ABPs 
   plotted versus time for
   different P\'{e}clet numbers at 
   $\eta = 0.29$. The dashed lines show the analytic results for free space from 
   Eqs.\ \eqref{eq:msd} and \eqref{eq:alpha_free}. The dashed-dotted line in the bottom graph indicates $\alpha = 2/z \approx 0.66$.
}
   \label{fig:msd_phio_03}
\end{figure}

In this section we 
study how 
ABPs move
in a two-dimen\-sional 
random
environment close to 
the
percolation
transition. Investigations of passive particles,
both diffusive and ballistic, demonstrate some universal behavior
\cite{Bauer2010}.
In Fig.\ \ref{fig:msd_phio_03}
we plot the mean squared displacement
$\langle \Delta r^2(t)\rangle$ and the 
local
exponent $\alpha(t)$ 
for ABPs moving
with different P\'eclet numbers in an environment with 
$\eta= 0.29$. 
For comparison, the
dashed 
lines 
show the analytic results 
for free space without any obstacles
from Eqs. \eqref{eq:msd} and \eqref{eq:alpha_free}.

Two features are visible. First, while in  free space the curves for the mean squared displacement are shifted upwards with
increasing $\mathrm{Pe}$, in the presence of the obstacles they roughly emanate from the same value at the smallest time 
$t=10^{-3} \tau_d$. The reason is that ABPs assemble at and swim against the obstacles as discussed in the beginning of 
Sec.\ \ref{sec:ABP}. So, in contrast to free space only a small fraction of them can move forward when rotational diffusion
orients their swimming directions away from the obstacles. In Sec.\ \ref{sec:VACF} we will rationalize this behavior by introducing an
effective swimming velocity $v_{\mathrm{eff}}$.

Second, while in free space the ABPs enter the diffusive regime for $t > \tau_d$, they all show subdiffusive motion in the
random environment close to the percolation transition. For intermediate times $ 1 <  t / \tau_d < 10^2$ and for all P\'eclet numbers
the exponent $\alpha$ stays close to the universal value $\alpha \approx 0.66$ shown by the dashed-dotted horizontal line in the lower plot.
This indicates that close to the percolation transition the dynamics of the ABP is entirely controlled by the random environment
regardless of $\mathrm{Pe}$. One can expect such a behavior from the results of Refs.\ \cite{Bauer2010,Spanner2016}. For the same
random environment they showed that in two dimensions the transport of passive particles, either Brownian or ballistic, 
shares the same universality class.

\section{Long-time diffusion and persistent motion}

\label{sec.same_random}

We now discuss two overall features of ABPs moving in a random Lorentz gas.
First, we look at the long-time diffusion coefficient for densities below the percolation transition and then address
how the persistent motion of an ABP is reduced in the presence of obstacles using the velocity autocorrelation function.

\subsection{Long-time diffusion coefficient}

\begin{figure}
\begin{center} 
	\resizebox{0.50\textwidth}{!}{%
	\includegraphics{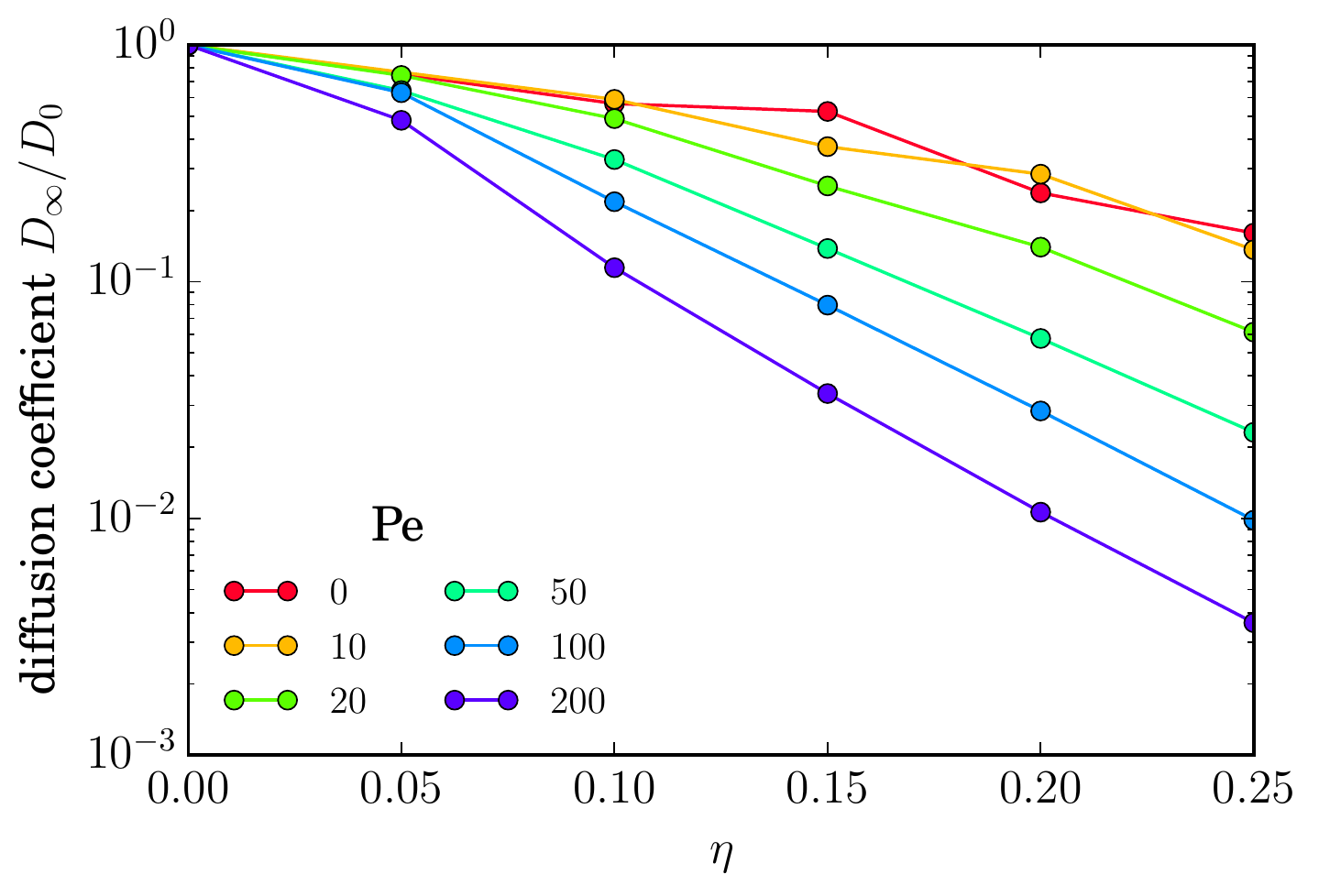}}
\end{center}
   \caption{Long-time diffusion coefficient $D_{\infty}$ versus $\eta$ for different $\mathrm{Pe}$. 
   $D_0$ is the coefficient for $\eta=0$.
}
   \label{fig:D_infty}
\end{figure}

Below the percolation threshold, the local diffusion constant $D(t)$, which we determine from the mean squared displacement
using Eq. \eqref{eq:diffusion}, assumes a constant value $D_\infty$ in the limit of long times.  In Fig.\ \ref{fig:D_infty} we plot 
$D_\infty$ versus $\eta$ for different $\mathrm{Pe}$ and refer it to the respective free-space value $D_0 (\eta =0)$.
All $D_{\infty}$ decrease with increasing obstacle density, however the decrease is stronger for larger P\'eclet numbers. 
While for passive particles the diffusion coefficient declines by less than an order of magnitude, it decreases by more than two
orders for $\mathrm{Pe} = 200$. The reason is that at such a high P\'eclet number the ABP is mostly stuck at the obstacles
and hardly explores the space in between.
This can be rationalized by an effective swimmining speed smaller than the free-space value $v_0$, which we introduce
in the following.

\subsection{Velocity autocorrelation function} 
\label{sec:VACF}

The velocity autocorrelation function (VACF) is a means to demonstrate the persistent motion of microswimmers. 
Therefore, we study now how the VACF changes for ABPs in the random Lorentz gas, in order to extract effective
values for swimming velocitiy, decorrelation time, and persistence length.

\paragraph{Free space}
In free space the velocity autocorrelation function 
$ C_0(t) = \langle \mathbf v (t) \cdot \mathbf v (0)\rangle $
follows
from the 
Langevin equation \eqref{eq:langevin}
setting $\mathbf{F}_{o}^{i} = \mathbf{0}$.
With $\dot{\mathbf r} = \mathbf v$ and the properties of Gaussian white noise as defined in  Eqs.\ \eqref{eq:noise_mean} and \eqref{eq:noise_corr}, 
the 
VACF
becomes \cite{Kapral2013}
\begin{align}
C_0(t) =\langle \mathbf v (t) \cdot \mathbf v (0)\rangle 
= v_0^2 \langle \mathbf e (t) \cdot \mathbf e (0)\rangle + 2 D^T \delta(t)
\end{align}
with the 
orientational
correlation function \cite{Dhont1996}
\begin{equation}
\langle \mathbf e(t)\cdot \mathbf e(0)\rangle = e^{-t/\tau_{r}} \, .  \label{eq:<e*e>}
\end{equation}
The
decorrelation time in two dimensions 
is
$\tau_r = 1/D^R$.
In the following we will also call it persistent time, since it gives the time scale on which the ABP moves persistently
in one direction.
Note, in the
absence of external forces, the deterministic velocity of the ABP 
always
points 
along
the orientation vector $\mathbf e(t)$, therefore $C_0(t)$ 
is determined by the autocorrelations 
of $\mathbf e(t)$.

\paragraph{Random environment}

 \begin{figure}
\begin{center} 
	\resizebox{0.50\textwidth}{!}{%
	\includegraphics{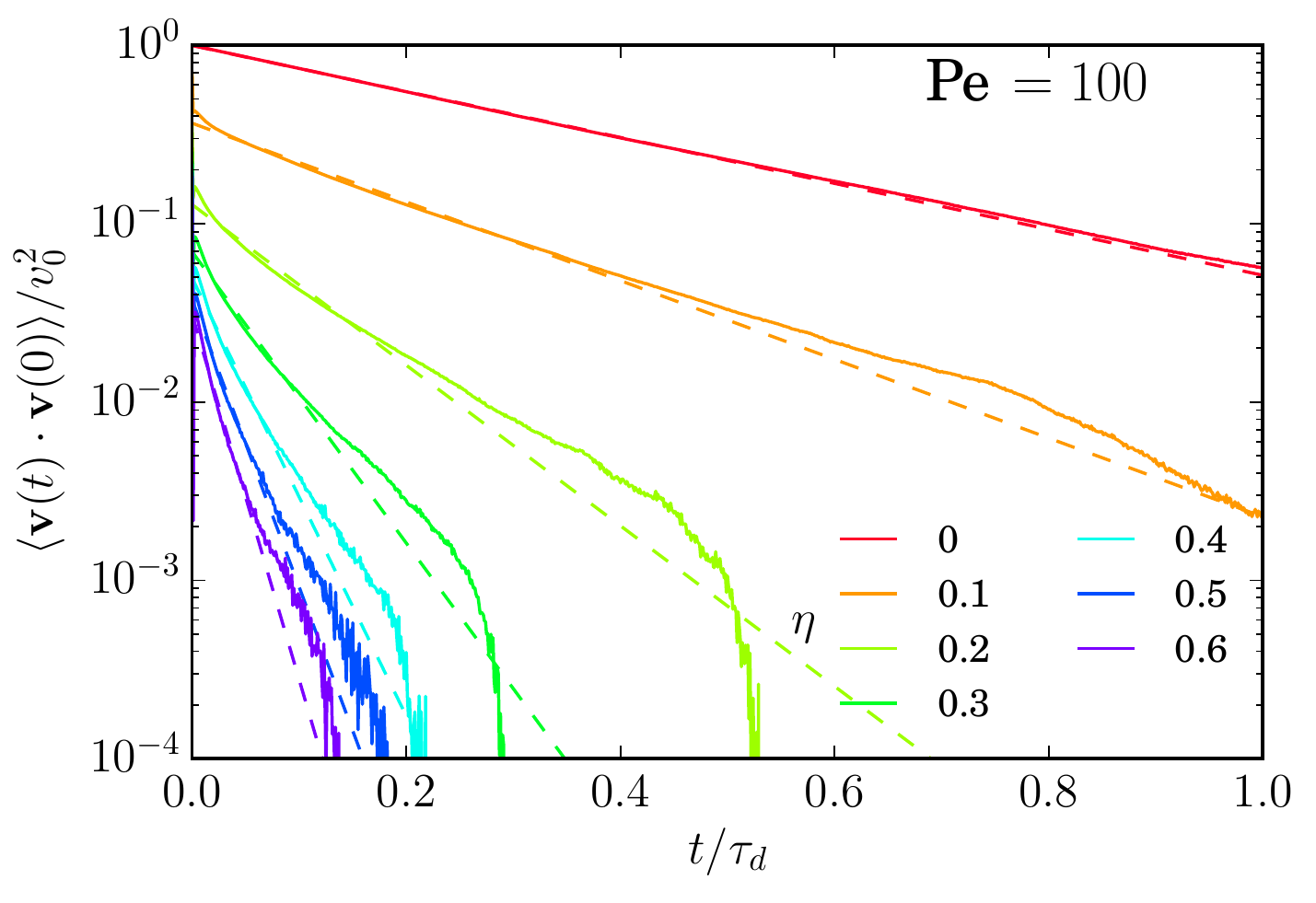}}
\end{center}
   \caption{Velocity autocorrelation functions for 
   $\mathrm{Pe}=100$ and different obstacle densities.
   The dashed lines 
   indicate the
   exponential fit
   from
   Eq. \eqref{eq:expfit}.}
   \label{fig:ac100log}
\end{figure}

In a
random
Lorentz gas 
an ABP
interacts sterically with the obstacles, 
where it cannot move along its orientation vector $\mathbf e(t)$.
It rather slides along the 
boundaries of the
obstacles 
or is
stuck in 
traps
 with convex shape formed by the obstacles. Both cases are illustrated
in Fig. \ref{fig:trajectory}.
Because 
we do not take into account
hydrodynamic interactions
with the obstacles and there are no other torques acting,
the dynamics of the orientation vector $\mathbf{e}(t)$ is not affected by the 
random
environment.
Thus,
its autocorrelation function follows Eq. \eqref{eq:<e*e>}. However, the full VACF strongly depends on the steric interactions of the ABP with the obstacles and
cannot 
be calculated analytically.
Therefore,
we determine
it from the swimmer trajectories.
We calculate 
velocity values at
discrete timesteps
[$\mathbf v(t) = \Delta \mathbf r (t)/\Delta t$],
numerically compute the Fourier transform of the time series, and then the VACF as
$C(t) = \int \langle |\mathbf{v}(\omega)|^2 \rangle  e^{i \omega t} d\omega/(2\pi) $ 
using the Wiener-Khinchin theorem \cite{Kubo1991}.

As an example, Fig.\ \ref{fig:ac100log} shows 
$C(t) = \langle \mathbf v (t) \cdot \mathbf v (0)\rangle$ 
in units of $v_02$
for $\mathrm{Pe}=100$ 
for different obstacle densities $\eta$.
While at $\eta = 0$ we nicely recover the expected exponential decay from the orientational decorrelation, we observe deviations from 
it with increasing $\eta$. The autocorrelations in velocity decay faster with increasing $\eta$ and then drop sharply to zero.
At times much smaller than $\tau_d$ we can even see anticorrelations in $C(t)$ (not shown). They arise at times 
$t\leq \tau_1= 4 \tau_d/\mathrm{Pe}^2$, where thermal translational diffusion dominates over active propulsion [see Eq.\ \eqref{eq:tau1}].
While stuck at an obstacle and oriented towards it, Brownian motion moves the ABP away from the obstacle and thereby causes
these anticorrelations.

Approximating the 
VACF by an exponential,
\begin{equation}
C(t)  = v_\text{eff}^2 e^{-t/\tau_\text{eff}}  \, ,
\label{eq:expfit}
\end{equation}
allows us to define an effective propulsion velocity $v_\text{eff}$ and an effective persistence time $\tau_\text{eff}$.
The velocity
$v_\text{eff}$ can be interpreted as 
the
mean velocity of 
an ABP
in the crowded environment. 
It is smaller than $v_0$, since the VACF also averages over ABPs, which are stuck at obstacles or in traps.
The exponential 
curves determined by least square fits for the VACF data in the range $0.01 \tau_d < t < \tau_d$
are shown as dashed lines in Fig. \ref{fig:ac100log}. 

\begin{figure}
\begin{center}
  \resizebox{0.5\textwidth}{!}{%
	\includegraphics{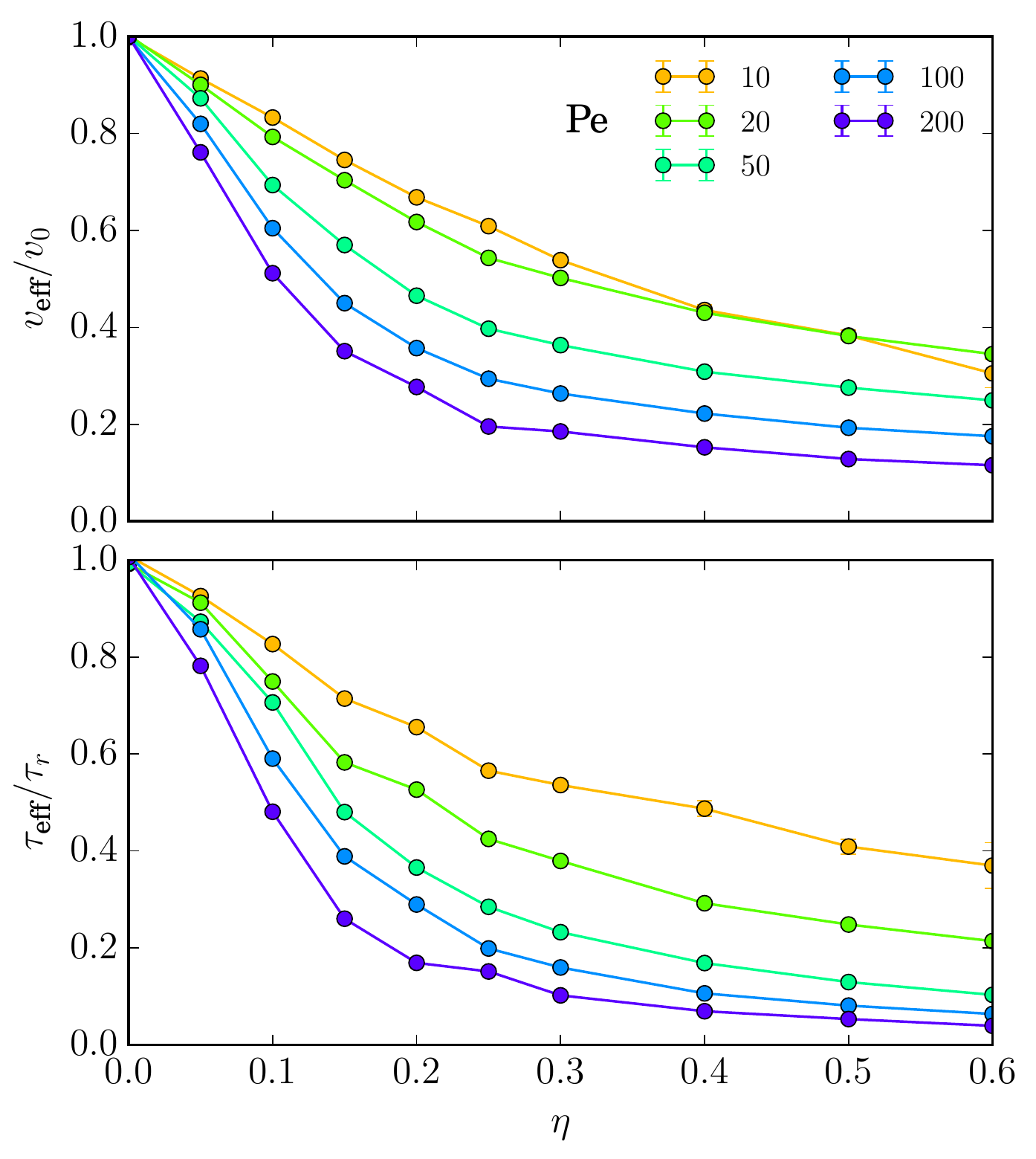}
		}
\end{center}
   \caption{Effective 
   propulsion
      velocity $v_\text{eff}$
      in units of $v_0$  
      (top) and effective 
      persistence
      time $\tau_\text{eff}$
      in units of $\tau_r$
      (bottom) 
     plotted versus      $\eta$ for different P\'eclet numbers. 
}
   \label{fig:v_tau_eff}
\end{figure}

In Fig. \ref{fig:v_tau_eff} we show the fit parameters $v_\text{eff}$ 
(top)
and $\tau_\text{eff}$
(bottom) rescaled, respectively, by the intrinsic propulsion velocity
$v_{0}$ and 
the orientational decorrelation time
$\tau_{r}$.
The effective velocity decreases with increasing $\eta$ since the ABP has less free space to move forward and thus the
probability to find it at an obstacle with zero or reduced velocity increases. The same is true for larger $\mathrm{Pe}$, where 
the ABP traverses the free space 
faster and thereby spends more time being stuck at the obstacles. The persistence time $\tau_\text{eff}$, on which velocity 
correlations decay, shows the same behavior. It decreases with increasing $\eta$ and $\mathrm{Pe}$, thus when the ABP 
spends more time at the obstacles, where its total velocity changes 
magnitude
and direction.

\begin{figure}
\begin{center}
  \resizebox{0.5\textwidth}{!}{%
	\includegraphics{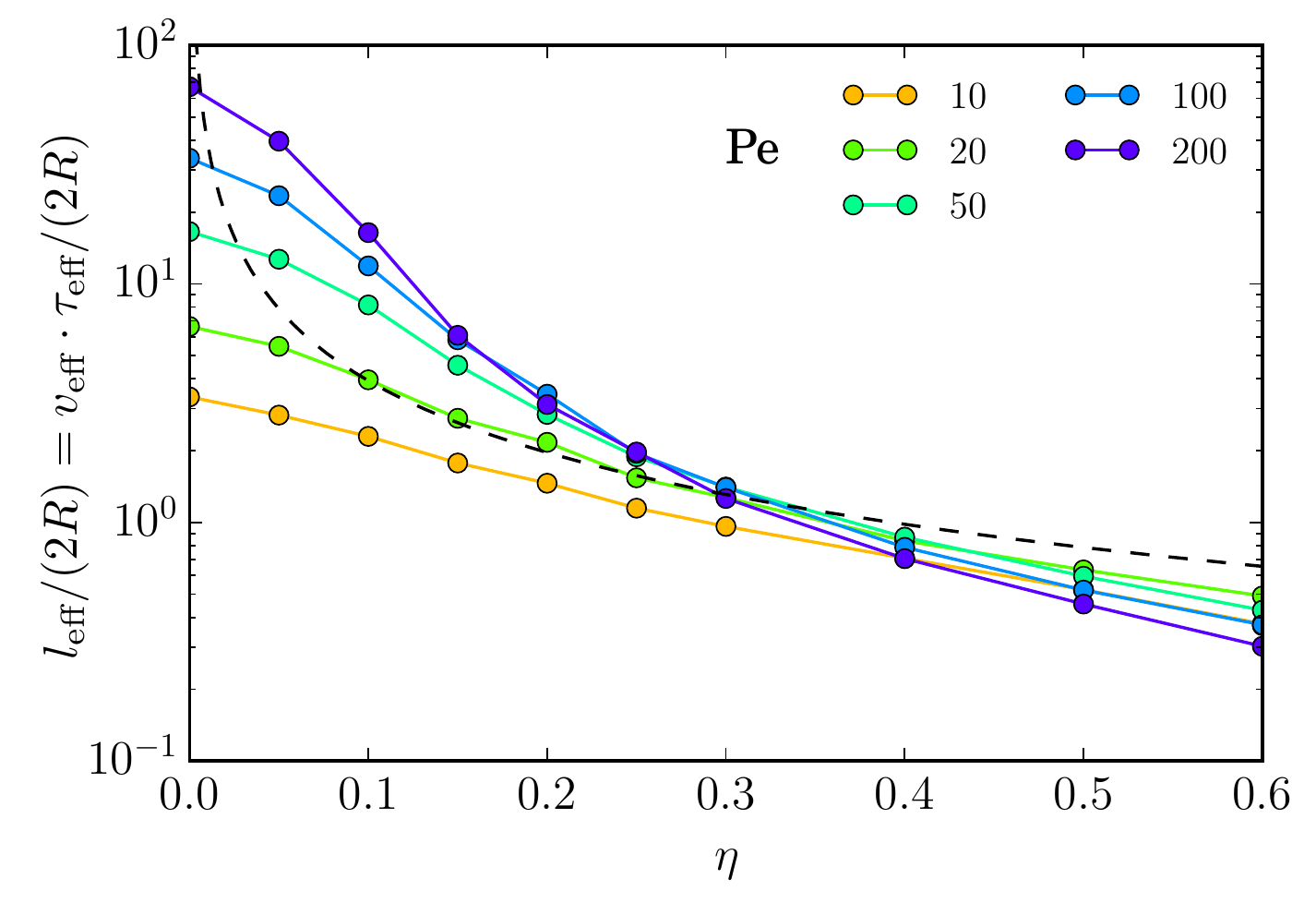}
		}
\end{center}
   \caption{Effective persistence length 
   $l_\text{eff} = v_\text{eff} \cdot \tau_\text{eff}$ in units of obstacle diameter $2R$ plotted versus $\eta$ for different $\mathrm{Pe}$.
  The black dashed line corresponds to the mean free
 path
  $l_\text{free}$ in a 
  random Lorentz gas.
}
     \label{fig:l_eff}
\end{figure}

It
is instructive to introduce an effective persistence length $l_\text{eff} = \tau_\text{eff} \cdot v_\text{eff}$,
which we plot in units of obstacle diameter $2R$
in Fig. \ref{fig:l_eff}. At $\eta=0$, 
$l_\text{eff}$ coincides of course with
the persistence length of an ABP in free space,
$v_0\tau_r$. It decreases with increasing $\eta$ but below $\eta = 0.3$ it rises with $\mathrm{Pe}$, which makes sense.
However, for sufficiently large $\eta$ we expect $l_{\text{eff}}$ to be governed by the random environment of the Lorentz gas,
for which the mean free path is $l_\text{free} = (R N / L^{2})^{-1} = \pi R/ 4\eta $ \cite{Torquato2002}. It is plotted as black dashed
line in Fig. \ref{fig:l_eff}.
Indeed, at around $\eta = 0.3$, \emph{i.e.,} above the percolation density $\eta_c\approx 0.28$, the effective persistence
lengths for different $\mathrm{Pe}$ are close to $l_\text{free}$, except for $\mathrm{Pe} = 10$, where translational Brownian motion
is still important. Above $\eta = 0.3$ they all lie below the geometrical length $l_\text{free}$ and are smallest for the largest $\mathrm{Pe}$,
as expected from the definition of $l_{\text{eff}}$.

\section{Summary and conclusions}
\label{sec:summary}

To conclude, in this article we studied the 
dynamics
of active Brownian particles in a two-dimensional heterogeneous environment of fixed obstacles modeled by a
random Lorentz gas.
Using Brownian dynamics simulations, we
explored how ABPs with different P\'eclet numbers move 
at
varying obstacle density. 
The percolation transition of the Lorentz gas plays a major role for
the
long-time dynamics of ABPs, separating long-range transport of ABPs on the one hand 
from
trapping of ABPs on the other hand.
At obstacle densities below the critical density, ABPs are 
able to diffuse over long distances, 
while at obstacle densities above the critical density all ABPs are trapped
in finite regions. This separation is independent of the P\'eclet number of the ABPs. 

Close to the critical obstacle density,
we observe that independent of the P\'eclet number,
ABPs show 
subdiffusive motion on intermediate time scales with the same
dynamic exponent 
as passive and ballistic particles.
Thus, we expect that in two dimensions they share the same universality class with passive diffusive and ballistic particles.
Therefore, 
the properties of the
random Lorentz gas alone control the transition between long-range transport and trapping of ABPs as well as the 
dynamic exponent at and close to 
the percolation transition.
However, in contrast to this universal behavior, we find that ABPs due to their persistent motion explore their 
environment faster than passive particles and,
consequently,
reach their long-time dynamics at earlier times. The persistent motion
also makes ABPs superdiffusive on intermediate times.

A main characteristic of ABPs is that they swim against obstacles and are stuck there until noise rotates them away from the
surface normal. As a result, the long-time diffusion constant decreases more strongly in a denser obstacle environment than 
for passive particles and this effect becomes stronger for large P\'eclet numbers. For example, for ABPs with $\mathrm{Pe} = 200$ 
the diffusion constant decreases by more than two orders of magnitude until an obstacle density of 0.25, while the decline for 
passive particles is below a factor of ten.

Diffusion of ABPs in free space is determined by propulsion velocity and persistence 
time, while in a random environment obstacles perturb the persistent motion of ABPs even at low densities. By measuring the 
velocity autocorrelation functions from the trajectories of the ABPs and making an exponential fit, we determined effective values 
for propulsion velocity and persistence time. They indeed decrease with increasing obstacle density and P\'eclet number and
thereby rationalize the observation for the long-time diffusion constant.

Based on the current study we aim at extending our investigations in different directions. For example, we will place the ABPs into
a random environment with a constant gradient in the obstacle density. We expect this setting to induce a drift of the ABPs towards the
denser region. One could view this as a form of taxis similar to motile cells exposed to  a substrate of varying stiffness.
The cells move along a gradient towards regions of largest substrate stiffness thus performing \emph{durotaxis} \cite{Lo2000,Plotnikov2013}.
ABPs also show a motility-induced phase separation \cite{Speck2014b,Cates2014,Takatori2015b,Blaschke2016},
where they phase-separate into a gas and dense phase for sufficiently large propulsion velocity and density. Immobile obstacles
can act as nucleation sites of dense ABP clusters\ \cite{Reichhardt2014} and we will explore how they influence the
phase behavior of ABPs.

\acknowledgement

We acknowledge helpful discussions with J. Blaschke, H. H. Boltz, F. H\"ofling, and R. Kruse and financial support from the Deutsche
Forschungsgemeinschaft in the framework of the collaborative research center SFB 910 and the research training group GRK 1558.
\bibliographystyle{myepjc}
\bibliography{library}

\end{document}